\documentclass[aps,prl,twocolumn,showpacs,superscriptaddress,showpacs]{revtex4-1}
\usepackage{xcolor}
\usepackage{physics}
\usepackage{graphicx}
\usepackage{dcolumn}
\usepackage{bm}

\begin{document}

\preprint{APS/123-QED}

\title{Weyl points created by a three-dimensional flat band}

\author{Yinong Zhou}
\affiliation{%
 Department of Materials Science and Engineering, University of Utah, Salt Lake City, Utah 84112, USA\\
 }%

\author{Kyung-Hwan Jin}
\affiliation{%
 Department of Materials Science and Engineering, University of Utah, Salt Lake City, Utah 84112, USA\\
 }%

\author{Huaqing Huang}
\affiliation{%
 Department of Materials Science and Engineering, University of Utah, Salt Lake City, Utah 84112, USA\\
 }%

\author{Zhengfei Wang}
\affiliation{%
Hefei National Laboratory for Physical Sciences at the Microscale,
CAS Key Laboratory of Strongly-Coupled Quantum Matter Physics,
University of Science and Technology of China, Hefei, Anhui 230026, China
}%

\author{Feng Liu}%
\email{fliu@eng.utah.edu}
\affiliation{%
 Department of Materials Science and Engineering, University of Utah, Salt Lake City, Utah 84112, USA\\
 }%
 \affiliation{%
 Collaborative Innovation Center of Quantum Matter, Beijing 100084, China\\
 }%

\date{\today}

\begin{abstract}
Following the discovery of topological insulators (TIs), topological Dirac/Weyl semimetal has attracted much recent interest. A prevailing mechanism for the formation of Weyl points is by breaking time reversal symmetry (TRS) or spatial inversion symmetry of a Dirac point. Here we demonstrate a generic formation mechanism for Weyl points by breaking TRS of a three-dimensional (3D) TI featured with highly degenerate 3D flat bands (FBs). It is in direct contrast to the conventional view that breaking TRS of a 2D/3D TI leads to a Chern insulator exhibiting quantum anomalous Hall effect. Based on a tight-binding model of pyrochlore lattice, we show that this unusual 3D-FB-enabled Weyl state may contain only a minimum of two Weyl points. Furthermore, using first-principles calculations, we identify this Weyl state in a real material Sn$_2$Nb$_2$O$_7$. The main features of the resulting Weyl points are analyzed with respect to symmetry, topological invariant and surface state. Our finding sheds new light on our fundamental understanding of topological physics and significantly extends the scope of Weyl semimetals to attract immediate experimental interest.
\end{abstract}

\maketitle


The mathematical concept of topology is an integral part to our fundamental understanding of a wide range of phenomena from high-energy particle physics to condensed-matter and materials physics. The discovery of topological insulators (TIs) \cite{kane2005quantum,bernevig2006quantum,konig2007quantum,zhou2014epitaxial,reis2017bismuthene} in solid materials has fostered a very active field of topological physics in the last decade. Following the study of TIs, Dirac/Weyl semimetals \cite{wan2011topological,xu2015discovery,lv2015observation,bergholtz2015topology} have attracted much interest. Fundamentally, a prevailing mechanism for the formation of Weyl points is by breaking time reversal symmetry (TRS) or inversion symmetry (INS) of a Dirac point \cite{xu2011chern,yang2015weyl,yan2017topological}. In this Rapid Communication, we demonstrate a generic mechanism for the formation of Weyl points by breaking TRS of a three-dimensional (3D) TI featured with highly degenerate 3D flat bands (FBs).
	
A TI phase can arise from either a semimetal \cite{kane2005quantum} or narrow-gap semiconductor \cite{bernevig2006quantum,konig2007quantum} through band inversion induced by spin-orbit coupling (SOC) that opens a global topological gap. Usually, breaking TRS of a TI leads to a Chern insulator (CI) exhibiting anomalous quantum Hall effect (AQHE). For example, the first experimental realization of AQHE \cite{chang2013experimental} is achieved by magnetic doping of a 3D TI film to break TRS \cite{yu2010quantized}. This TI-to-CI transition is general to both two-dimensional (2D) and 3D TIs. In contrast, however, we discover a TI-to-Weyl semimetal transition induced by breaking TRS of a 3D TI featured with FBs, while breaking TRS of a 2D TI featured with a FB would lead still to a CI \cite{wang2013quantum}.
	
A topological FB arises from destructive interference (phase cancellation) of Bloch wave functions, independent of the single-particle crystalline Hamiltonian (i.e., the strength of lattice hopping) \cite{zheng2014exotic}. The completely quenched electronic kinetic energy in a FB magnifies any finite electron-electron interaction, leading to a range of exotic quantum phases, such as ferromagnetism \cite{mielke1991ferromagnetism,mielke1991ferromagnetic,mielke1992exact,tasaki1992ferromagnetism,zhang2010proposed}, Wigner crystallization \cite{wu2007flat,wu2008p}, and superconductivity \cite{miyahara2007bcs,kobayashi2016superconductivity}. Also, the phase cancellation renders the FB to be inherently topologically nontrivial, leading to high-temperature fractional QHE \cite{tang2011high,neupert2011fractional}. Many quantum phases exist commonly for both 2D and 3D FBs, such as ferromagnetic \cite{kimura2002magnetic,tanaka2003stability,pollmann2008kinetic,hase2018possibility} and TI phase \cite{liu2009spin,guo2009three,guo2009topological,wang2010quantum,kurita2011topological,hatsugai2011zq,sun2011nearly,wang2013prediction}. Interestingly, our finding represents another unique manifestation of 3D topological FBs in forming Weyl semimetal phase, which is absent for 2D FB.
	
Using pyrochlore lattice as a prototypical example, we first illustrate this unusual topological phase transition from a 3D TI to a Weyl semimetal using tight-binding (TB) model analysis considering two extreme cases: large magnetization vs large SOC. We show that the Weyl semimetal so formed may contain only a minimum of two Weyl points when the band splitting induced by magnetization (breaking TRS) is much larger than the SOC gap of the intermediate TI phase, while four Weyl points will be formed when the former is smaller than the latter. Then we further reveal its physical manifestations in a real material of Sn$_2$Nb$_2$O$_7$, using first-principles calculations. The main features of the resulting Weyl points are analyzed with respect to symmetry, topological invariant, and surface state. The details of our calculation method are presented in the Supplemental Material \footnote{See Supplemental Materials at http://link.aps.org/supplemental/xxx, for more details which include Ref.~\cite{blochl1994projector,kresse1999ultrasoft,perdew1996generalized,perdew1996generalized,kresse1996efficient,methfessel1989high,hobbs2000fully,marzari1997maximally,wu2018wanniertools}.}.

\begin{figure}
	\includegraphics[width=1\columnwidth]{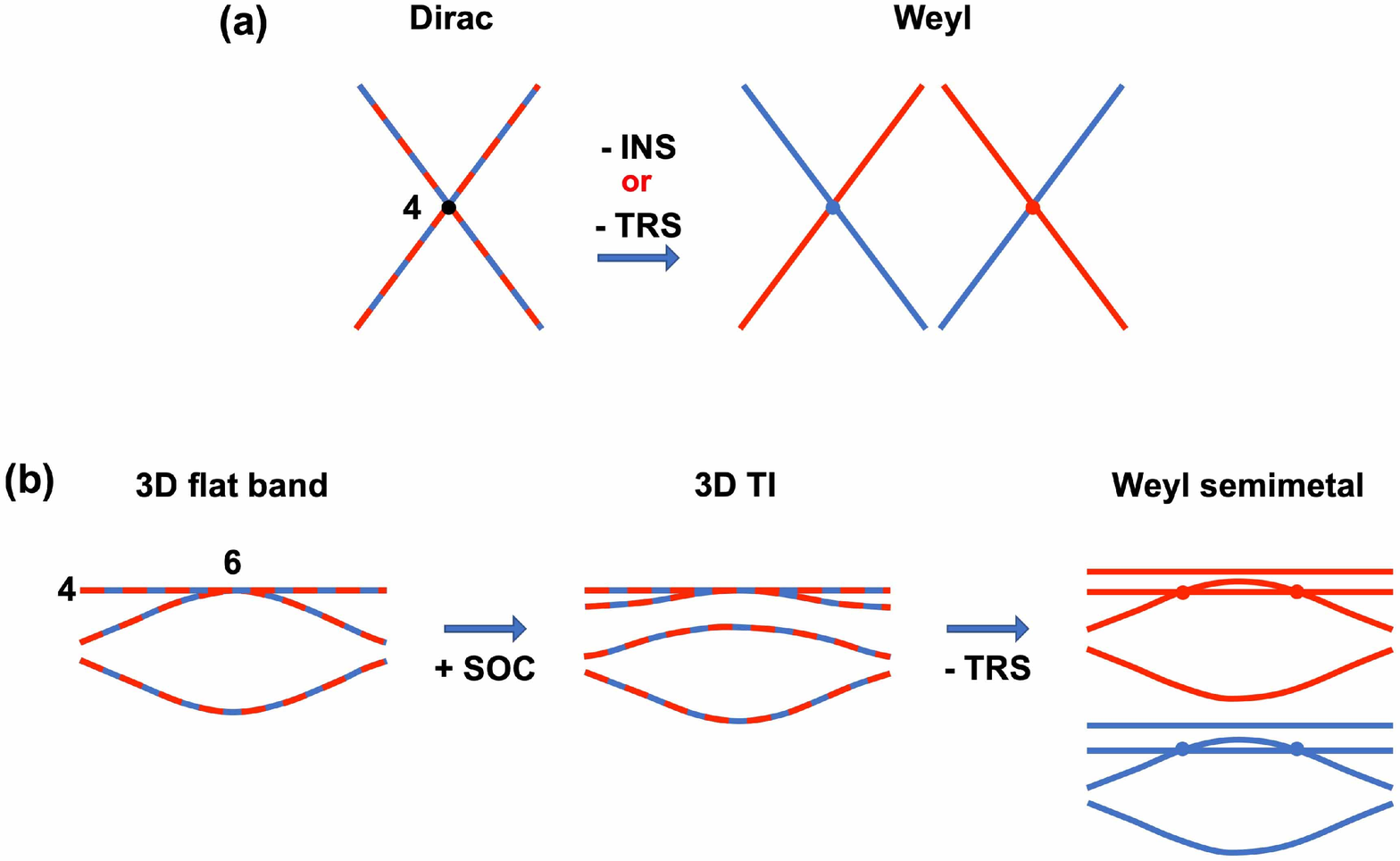}
	\caption{\label{fig:epsart} Schematic illustration of a conventional (a) and a new mechanism (b) of Weyl points creation. (a) From a four-fold degenerate Dirac point to a pair of Weyl points by breaking TRS or INS. (b) From a 3D four-fold degenerate flat band and a six-fold degenerate touching point at the $\Gamma$ point to a 3D TI with SOC interaction, then to a Weyl semimetal by breaking TRS. A pair of Weyl points are created for each spin branch.}
\end{figure}

We begin with a general introduction of this unconventional mechanism for the creation of Weyl points, i.e., the Weyl points formed by breaking TRS of 3D FBs in the presence of SOC. As shown in Fig. 1, the Dirac point is a four-fold degenerate point protected by TRS and INS. To generate a pair of doubly degenerate Weyl points from a Dirac point one may break either TRS or INS [Fig. 1(a)] \cite{xu2011chern,yang2015weyl,yan2017topological}. On the other hand, the 3D FB in a pyrochlore lattice has a four-fold degeneracy including spin [left panel of Fig. 1(b)]. Thus, the touching point of the FB and dispersive band at the $\Gamma$ point is a six-fold degenerate point. If only TRS were broken, this six-fold degenerate point would split into two three-fold degenerate points in a ferromagnetic state \cite{hase2018possibility} and no Weyl point would be formed. However, in the presence of SOC, the six-fold degenerate point will split into a four-fold degenerate point and a doubly degenerate point, as a non-trivial band gap is opened between the FB and the dispersive band, leading to a 3D TI phase [middle panel of Fig. 1(b)] \cite{guo2009three}. Then, if one further lifts the degeneracy by breaking TRS, a pair of Weyl points will be generated due to the crossing of the FB and the dispersive band [right panel of Fig. 1(b)]. Thus, in the presence of both SOC and Zeeman field, the 3D FB system becomes a Weyl semimetal. Furthermore, for a sufficiently large band splitting, i.e. the magnetization strength being much stronger than the SOC strength, there will be only minimum two Weyl points present, providing an ideal Weyl semimetal phase for experimental characterization. Note that for 2D FB systems, in presence of SOC, breaking TRS would instead lead to a well-known transition from 2D TI to CI (Fig. S1) \cite{wang2013quantum}.

Next, we elaborate further on the above concept by a rigorous TB model analysis on a pyrochlore lattice. The pyrochlore lattice has a face-centered-cubic structure with the group symmetry $\textit{Fd}\bar{3}\textit{m}$ \cite{40footnote}. To describe the band dispersion and topological properties of the pyrochlore lattice, we construct a single-orbital nearest-neighbor (NN) hopping TB model with the following Hamiltonian:

\begin{multline}
H=-t\sum_{\langle{ij}\rangle\alpha}{c_{i\alpha}^\dagger}c_{j\alpha}\\
+i\lambda_{soc}\sum_{\langle\langle{ij}\rangle\rangle\alpha\beta}\frac{2}{\sqrt{3}}(\hat{\bf r}_{ij}^1\times\hat{\bf r}_{ij}^2)\cdot{\bf\sigma}_{\alpha\beta}{c_{i\alpha}^\dagger}c_{j\beta} \\
+\lambda_z\sum_{i\alpha}{c_{i\alpha}^\dagger}{\bf\sigma}_zc_{i\alpha}.
\end{multline}
The first term represents the NN hopping, c$_{i\alpha}^\dagger$  (c$_{i\alpha}$)  is the operator of the electron creation (annihilation) on site $\textit{i}$ of spin $\alpha$. The second term represents the SOC with the coupling strength $\lambda_{soc}$ \cite{guo2009three}, where unit vectors $\hat{\bf r}_{ij}^{1,2}$ are indicated in Fig. 2(a) on the upper right tetrahedron; $\frac{2}{\sqrt{3}}$$\vert\hat{\bf r}_{ij}^1\times\hat{\bf r}_{ij}^2\vert$ = $\pm$1; and $\bf\sigma$ is the Pauli matrix. The third term represents the Zeeman splitting with the exchange field strength $\lambda_z$.

Figure 2 shows the TB band structures of pyrochlore lattice [see Fig. 2(a)]. Fermi level is set at E=0, considering a half-filled system. Without SOC and magnetization, four bands [Fig. 2(c)] are drawn along the high symmetry points as indicated in Fig 2(b); the top two bands are flat and degenerate with each other without SOC. These four bands arise from four $\textit{s}$ orbitals (atomic basis) sitting at the four corners of a tetrahedron in the unit cell. Effectively, one may view the tetrahedron as a ``molecule'' with a molecular $\textit{sp}^3$ orbital basis, and from parity analysis at $\Gamma$ point (Fig. S4), one can assign the bottom band as the molecular $\textit{s}$-orbital band, the middle dispersive band as the $\textit{p}_z$-orbital band, and the top two degenerate FBs as ($\textit{p}_x$, $\textit{p}_y$)-orbital bands, as indicated in Fig. 2(c). The six-fold degenerate point at the $\Gamma$ is thus due to the degeneracy of three $\textit{p}$ orbitals. We note that our pyrochlore lattice model without considering correlation has a band order of 4-2-2 degeneracy which agrees with the DFT band structure of Sn$_2$Nb$_2$O$_7$ \cite{hase2018possibility,guo2009three} (see below), but different from that of 2-4-2 in Y$_2$Ir$_2$O$_7$ \cite{wan2011topological} which has a different Weyl semimetal phase due to intermediate correlation and strong SOC.

\begin{figure}
\includegraphics[width=1\columnwidth]{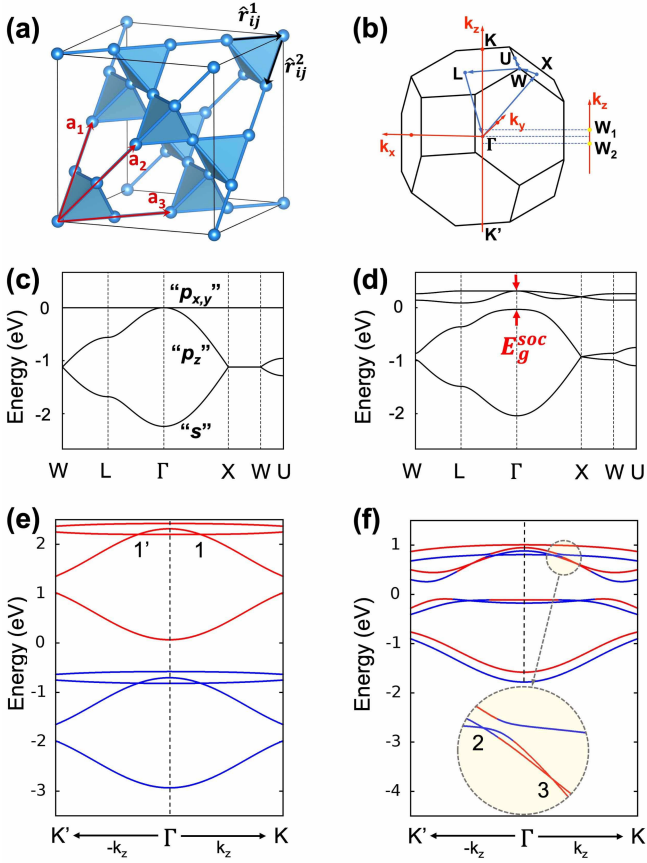}
\caption{\label{fig:epsart} (a) The pyrochlore lattice structure. The vectors $\hat{\bf r}_{ij}^{1,2}$ for calculating SOC in Eq. (1) are indicated on the upper right tetrahedron. (b) The first Brillouin zone of pyrochlore lattice with high-symmetry lines and points indicated; W$_{1,2}$ are two positions along the $\textbf{\textit{k}}_z$ direction close to the Weyl points (when magnetization direction is along $\textbf{\textit{k}}_z$). (c),(d) TB band structures (c) without SOC and magnetization, the equivalent molecular $\textit{sp}^3$ orbital bases are labeled on the corresponding bands; (d) with SOC but without magnetization. The SOC band gap at $\Gamma$ is labeled as $\textit{E}_g^{soc}$. (e),(f) TB band structures along the $\textbf{\textit{k}}_z$ direction for two extreme cases: (e) $\lambda_z\gg\lambda_{soc}$ and (f) $\lambda_z\ll\lambda_{soc}$. The Weyl points are numbered as 1-3; their counterparts on the opposite direction are labeled as 1'-3'. The blue (red) color represents spin-up (-down) bands.}
\end{figure}

Next, we turn to SOC, in which a topological nontrivial band gap is opened at the $\Gamma$ point between the top FBs and the middle dispersive band [Fig. 2(d)]. The two FBs are also gapped except at the high-symmetry points $\Gamma$ and X. Moreover, if we turn on magnetization, the system becomes ferromagnetic. If the band splitting is much larger than the SOC gap ($\lambda_z\gg\lambda_{soc}$), then the spin-up and -down bands are fully split. The corresponding band structure along $\textbf{\textit{k}}_z$ is shown in Fig. 2(e). There appears a band crossing between the lower FB and the middle dispersive band to form two Weyl points (1 and 1'), which is the minimum number of Weyl points that can exist in a system with inversion symmetry. On the other hand, when magnetization is much smaller than the SOC ($\lambda_z\ll\lambda_{soc}$), spin-up and -down bands are mixed with each other. Four Weyl points are generated by the crossings between the split FBs, as shown in Fig. 2(f). Note that with SOC, there are four-fold degenerate points at the X point, as shown in Fig. 2(d). But upon magnetization, unlike the Weyl points formation around the $\Gamma$ point, two-fold degenerate lines are created along X-W high-symmetry path (Fig. S5). Also, the z-component of spin ($\textit{S}_z$) is generally not conserved in the presence of SOC. However, when $\lambda_z\gg\lambda_{soc}$, $\textit{S}_z$ is approximately conserved as shown in Fig. 2(e).

The transition from four Weyl points with $\lambda_z\ll\lambda_{soc}$ to two Weyl points with $\lambda_z\gg\lambda_{soc}$ is understandable, because in the presence of SOC, the original perfect FBs become partially dispersive being gapped everywhere except at $\Gamma$ and X. A small exchange splitting will induce crossings of the four FBs to form four Weyl points [Fig. 2(f)]. In contrast, a strong exchange splitting will completely separate spin-up and -down bands. Then the SOC gapping will cause a crossing between the lower FB and the middle dispersive band forming two Weyl points [Fig. 2(e)]. The larger the $\lambda_{soc}$, the larger the $\lambda_z$ is needed to induce the transition. To show this in more detail, one can convert $\lambda_z$ into the exchange energy splitting $\Delta$\textit{E}$^z$ and $\lambda_{soc}$ into the SOC gap $\textit{E}_g^{soc}$. It can be shown by solving Eq. (1) that $\Delta$\textit{E}$^z$ = 2$\lambda_z$ and $\textit{E}_g^{soc}$ = 8${\sqrt{3}}\lambda_{soc}$. For a given $\textit{E}_g^{soc}$, from the plot of distribution of Weyl points as a function of the ratio $\Delta$\textit{E}$^z$/$\textit{E}_g^{soc}$ (Fig. S6), one can extract the transition point from four Weyl points to two Weyl points, which indicates $\Delta$\textit{E}$^z$ increases linearly with $\textit{E}_g^{soc}$ with a coefficient of $\sim$ 0.36 (see inset of Fig. S6).

In the regime of $\lambda_z\ll\lambda_{soc}$, the four Weyl points move closer with the increase of exchange field, until they annihilate each other at a low-symmetry $\textit{k}$-point. At the same time, two new Weyl points are generated at the Brillouin zone (BZ) boundary entering the regime of $\lambda_z\gg\lambda_{soc}$. These two Weyl points will move closer with the further increase of the exchange field and then stay fixed for sufficiently large Zeeman field when the spin-up and -down bands are totally separated. The location of Weyl points is found to be always along the magnetization direction (see, e.g., the case of two Weyl points in Fig. S7).

 \begin{figure}
 	\includegraphics[width=1\columnwidth]{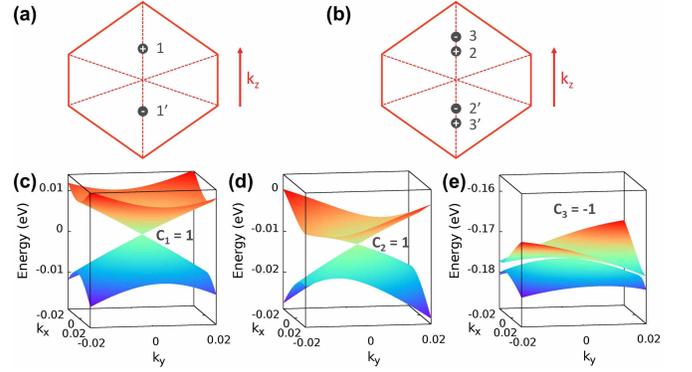}
 	\caption{\label{fig:epsart} (a),(b) Position of Weyl points for (a) $\lambda_z\gg\lambda_{soc}$ and (b) $\lambda_z\ll\lambda_{soc}$. (c)-(e) 3D band structure and Chern number of Weyl points in the $\textbf{\textit{k}}_x$-$\textbf{\textit{k}}_y$ plane for (c) point 1, (d) point 2, and (e) point 3.}
 \end{figure}

To confirm the topology of the Weyl points, Chern number (C) is calculated using the Kubo formula \cite{thouless1982quantized}
\begin{equation}
C=\frac{1}{2\pi}\int_\textit{BZ}d^2\textbf{\textit{k}}\bf\Omega(\textbf{\textit{k}}),\nonumber
\end{equation}
\begin{equation}
\bf\Omega(\textbf{\textit{k}})=-\textit{Im}\sum_{\textit{m}\neq\textit{n}}\frac{
	\bra{\Psi_{\textit{n},\textbf{\textit{k}}}}\nabla_{\textbf{\textit{k}}}\textit{H}_\textbf{\textit{k}}\ket{\Psi_{\textit{m},\textbf{\textit{k}}}}
	\times\bra{\Psi_{\textit{m},\textbf{\textit{k}}}}\nabla_{\textbf{\textit{k}}}\textit{H}_\textbf{\textit{k}}\ket{\Psi_{\textit{n},\textbf{\textit{k}}}}}
{(\textit{E}_{\textit{m},\textbf{\textit{k}}}-\textit{E}_{\textit{n},\textbf{\textit{k}}})^2},
\end{equation}
where \textit{m}, \textit{n} are band indexes; $\ket{\Psi_{\textit{m},\textbf{\textit{k}}}}$, $\ket{\Psi_{\textit{n},\textbf{\textit{k}}}}$ are eigenstates; \textit{E}$_\textit{m}$,\textit{E}$_\textit{n}$ are eigenvalues; and $\bf\Omega$ is the Berry phase connection of the Bloch state. Figure 3(a) and 3(b) show the position and chirality of Weyl points for the two cases, $\lambda_z\gg\lambda_{soc}$ and $\lambda_z\ll\lambda_{soc}$, respectively. Using Eq. (2), the Chern number is calculated on a 2D manifold enclosing the Weyl point. For points 1 and 2, C = 1; for point 3, C = -1. The points 1', 2', and 3' have the opposite Chern number to points 1, 2, and 3, respectively. The non-zero Chern numbers are also confirmed by plotting the flux of Berry connections around these band crossing points (Fig. S8). Therefore, we confirm they are indeed Weyl points.

\begin{figure}
	\includegraphics[width=1\columnwidth]{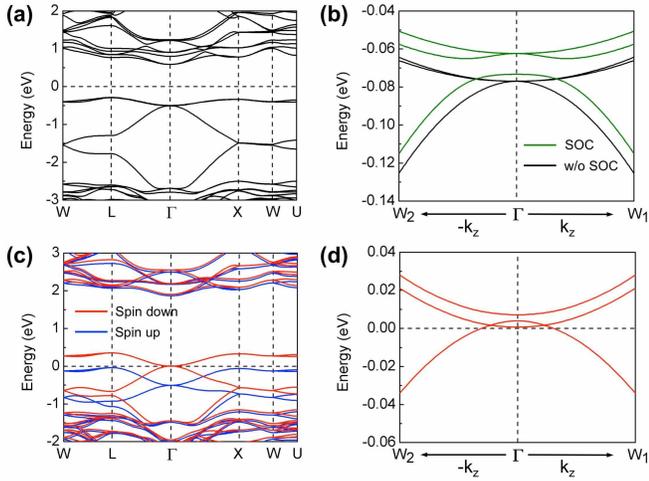}
	\caption{\label{fig:epsart} DFT band structure of Sn$_2$Nb$_2$O$_7$: (a) without SOC and magnetization, the four bands below Fermi level are mainly from Sn-s orbitals; (c) with magnetization and SOC, spin-up (blue) and spin-down (red) bands split; (b) and (d) are the zoom-in plots of (a) and (c), respectively; in (b), green (black) bands are results with (without) SOC; in (d) only bands of one spin component are shown to better illustrate the two Weyl points formed along the magnetization direction.}
\end{figure}

Lastly, we demonstrate a real material system to realize the above-mentioned unexpected Weyl semimetal associated with the 3D FB. The 3D FB material should meet the following conditions: (\textit{i}) the bands formed by the atoms on the tetrahedral motif are isolated from other bands; (\textit{ii}) the system has a considerable SOC strength; (\textit{iii}) the system can be magnetized. Specifically, we narrow down our search to pyrochlore oxide, which has a common formula A$_2$B$_2$O$_7$, where A is either a trivalent rare earth ion or a mono/divalent cation, and B is a transition metal or a \textit{p}-block metal ion. One well-known pyrochlore oxide is Sn$_2$Nb$_2$O$_7$. It contains Sn atoms forming the corner-shared tetrahedral motifs, whose \textit{s}-orbitals constitute the desired single-orbital TB model in a pyrochlore lattice. Importantly, Sn$_2$Nb$_2$O$_7$ has been shown to exhibit a TI phase in the presence of SOC without magnetization \cite{guo2009three} and a ferromagnetic instability upon doping of its 3D FB without SOC \cite{hase2018possibility}. Therefore, it would be very interesting to find out what happens when both SOC and doping are considered.

Using density functional theory (DFT), we calculate the band structures of Sn$_2$Nb$_2$O$_7$. Four ``pyrochlore” bands below Fermi level arising from the Sn-\textit{s} orbitals are confirmed to be isolated from other bands, as shown in Fig. 4(a). With the SOC, an 11 meV band gap opens between FBs and dispersive bands below at $\Gamma$, as shown by the magnified view of green bands in Fig 4b. The system is confirmed to be a 3D TI with Z$_2$ = 1 \cite{kane2005z}, in agreement with the previous study \cite{guo2009three}. On the other hand, without SOC, because the quenched kinetic energy of FB magnifies the Coulomb interaction, a partially filled FB will exhibit a ferromagnetic instability with hole doping \cite{hase2018possibility}. As shown in Fig. 4(c), the spin-up and -down bands are split upon doping of two holes. This is the largest exchange splitting with the highest magnetic momentum of 2$\mu_B$ per unit cell, which decreases with the decrease of doping concentration. Then, considering the SOC with doping [Fig. 4(d)], two inversion-symmetric Weyl points are generated at ($\pm$0.0115, $\pm$0.0115, $\pm$0.023) in the $\textit{k}$-space along the magnetization direction $\textbf{\textit{k}}_z$. Apparently, this specific Weyl semimetal phase corresponds to the regime of $\lambda_z\gg\lambda_{soc}$ described by the TB model above in Fig. 2(e).

To confirm the topology of Weyl points, we calculated the Chern number using Wannier functions \cite{mostofi2008wannier90}. When the $\textbf{\textit{k}}_x$-$\textbf{\textit{k}}_y$ plane crosses the two Weyl points from bottom to top, the Chern number jumps first from 0 to -1 and then from -1 to 0 [Fig. 5(a)], signifying the topology of the Weyl points. Another significant topological manifestation of Weyl point is the emergence of the Fermi arcs and surface states, which are shown in Fig. 5(b) and 5c respectively. One clearly sees a Fermi arc connecting two Weyl points with opposite chirality in Fig. 5(b) and surface states along $\textbf{\textit{k}}_z$ in Fig. 5(c); both are on the (100) surface.

\begin{figure}
	\includegraphics[width=1\columnwidth]{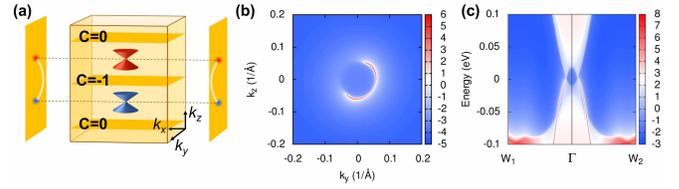}
	\caption{\label{fig:epsart} (a) Schematics of a Weyl semimetal with Chern numbers and Fermi arcs indicated. (b) The Fermi arc on the (100) surface of Sn$_2$Nb$_2$O$_7$ at $\textit{E}$ = 0.001 eV. (c) Bulk band structure (light shaded regions) and surface states on the (100) surface (dark lines) of Sn$_2$Nb$_2$O$_7$.}
\end{figure}

The DFT results can be well fitted by a TB model with the following parameters: on-site energy of -0.756 eV, hopping parameter of -0.28 eV, $\lambda_{soc}$ = 0.75$\times$10$^{-3}$ eV and $\lambda_z$ = 0.2 eV. In addition, the DFT calculations also confirm that with a small doping concentration, four Weyl points can be formed corresponding to the regime of $\lambda_z\ll\lambda_{soc}$ as described by the TB model. The existence of Weyl semimetal phase at different doping concentrations is beneficial for experimental measurement because some range of doping might be achieved more easily.

In conclusion, we revealed a mechanism of 3D-FB-enabled creation of Weyl points, in contrast with the conventional mechanism of lowering the symmetry of a Dirac point. This mechanism is exemplified in a real material of Sn$_2$Nb$_2$O$_7$, which can be potentially extended to other 3D FB materials. In addition, we have demonstrated a fundamental difference between the 2D and 3D FBs with regard to their associated topological phases, which is rooted in their distinct dimensions and band degeneracies. Our findings significantly enrich the topological physics associated with Weyl points and FBs; and provide a promising platform to extend the search for new Weyl semimetal materials.

This work is supported by U.S. DOE-BES (Grant No. DE-FG02-04ER46148). The calculations were done on the CHPC at the University of Utah and DOE-NERSC.
Y. Z. and K.-H. J. contributed equally to this work.

\nocite{*}


\begin{thebibliography}{52}%
	\makeatletter
	\providecommand \@ifxundefined [1]{%
		\@ifx{#1\undefined}
	}%
	\providecommand \@ifnum [1]{%
		\ifnum #1\expandafter \@firstoftwo
		\else \expandafter \@secondoftwo
		\fi
	}%
	\providecommand \@ifx [1]{%
		\ifx #1\expandafter \@firstoftwo
		\else \expandafter \@secondoftwo
		\fi
	}%
	\providecommand \natexlab [1]{#1}%
	\providecommand \enquote  [1]{``#1''}%
	\providecommand \bibnamefont  [1]{#1}%
	\providecommand \bibfnamefont [1]{#1}%
	\providecommand \citenamefont [1]{#1}%
	\providecommand \href@noop [0]{\@secondoftwo}%
	\providecommand \href [0]{\begingroup \@sanitize@url \@href}%
	\providecommand \@href[1]{\@@startlink{#1}\@@href}%
	\providecommand \@@href[1]{\endgroup#1\@@endlink}%
	\providecommand \@sanitize@url [0]{\catcode `\\12\catcode `\$12\catcode
		`\&12\catcode `\#12\catcode `\^12\catcode `\_12\catcode `\%12\relax}%
	\providecommand \@@startlink[1]{}%
	\providecommand \@@endlink[0]{}%
	\providecommand \url  [0]{\begingroup\@sanitize@url \@url }%
	\providecommand \@url [1]{\endgroup\@href {#1}{\urlprefix }}%
	\providecommand \urlprefix  [0]{URL }%
	\providecommand \Eprint [0]{\href }%
	\providecommand \doibase [0]{http://dx.doi.org/}%
	\providecommand \selectlanguage [0]{\@gobble}%
	\providecommand \bibinfo  [0]{\@secondoftwo}%
	\providecommand \bibfield  [0]{\@secondoftwo}%
	\providecommand \translation [1]{[#1]}%
	\providecommand \BibitemOpen [0]{}%
	\providecommand \bibitemStop [0]{}%
	\providecommand \bibitemNoStop [0]{.\EOS\space}%
	\providecommand \EOS [0]{\spacefactor3000\relax}%
	\providecommand \BibitemShut  [1]{\csname bibitem#1\endcsname}%
	\let\auto@bib@innerbib\@empty
	\bibitem [{\citenamefont {Kane}\ and\ \citenamefont
		{Mele}(2005{\natexlab{a}})}]{kane2005quantum}%
	\BibitemOpen
	\bibfield  {author} {\bibinfo {author} {\bibfnamefont {C.~L.}\ \bibnamefont
			{Kane}}\ and\ \bibinfo {author} {\bibfnamefont {E.~J.}\ \bibnamefont
			{Mele}},\ }\href@noop {} {\bibfield  {journal} {\bibinfo  {journal} {Physical
				review letters}\ }\textbf {\bibinfo {volume} {95}},\ \bibinfo {pages}
		{226801} (\bibinfo {year} {2005}{\natexlab{a}})}\BibitemShut {NoStop}%
	\bibitem [{\citenamefont {Bernevig}\ \emph {et~al.}(2006)\citenamefont
		{Bernevig}, \citenamefont {Hughes},\ and\ \citenamefont
		{Zhang}}]{bernevig2006quantum}%
	\BibitemOpen
	\bibfield  {author} {\bibinfo {author} {\bibfnamefont {B.~A.}\ \bibnamefont
			{Bernevig}}, \bibinfo {author} {\bibfnamefont {T.~L.}\ \bibnamefont
			{Hughes}}, \ and\ \bibinfo {author} {\bibfnamefont {S.-C.}\ \bibnamefont
			{Zhang}},\ }\href@noop {} {\bibfield  {journal} {\bibinfo  {journal}
			{Science}\ }\textbf {\bibinfo {volume} {314}},\ \bibinfo {pages} {1757}
		(\bibinfo {year} {2006})}\BibitemShut {NoStop}%
	\bibitem [{\citenamefont {K{\"o}nig}\ \emph {et~al.}(2007)\citenamefont
		{K{\"o}nig}, \citenamefont {Wiedmann}, \citenamefont {Br{\"u}ne},
		\citenamefont {Roth}, \citenamefont {Buhmann}, \citenamefont {Molenkamp},
		\citenamefont {Qi},\ and\ \citenamefont {Zhang}}]{konig2007quantum}%
	\BibitemOpen
	\bibfield  {author} {\bibinfo {author} {\bibfnamefont {M.}~\bibnamefont
			{K{\"o}nig}}, \bibinfo {author} {\bibfnamefont {S.}~\bibnamefont {Wiedmann}},
		\bibinfo {author} {\bibfnamefont {C.}~\bibnamefont {Br{\"u}ne}}, \bibinfo
		{author} {\bibfnamefont {A.}~\bibnamefont {Roth}}, \bibinfo {author}
		{\bibfnamefont {H.}~\bibnamefont {Buhmann}}, \bibinfo {author} {\bibfnamefont
			{L.~W.}\ \bibnamefont {Molenkamp}}, \bibinfo {author} {\bibfnamefont {X.-L.}\
			\bibnamefont {Qi}}, \ and\ \bibinfo {author} {\bibfnamefont {S.-C.}\
			\bibnamefont {Zhang}},\ }\href@noop {} {\bibfield  {journal} {\bibinfo
			{journal} {Science}\ }\textbf {\bibinfo {volume} {318}},\ \bibinfo {pages}
		{766} (\bibinfo {year} {2007})}\BibitemShut {NoStop}%
	\bibitem [{\citenamefont {Zhou}\ \emph {et~al.}(2014)\citenamefont {Zhou},
		\citenamefont {Ming}, \citenamefont {Liu}, \citenamefont {Wang},
		\citenamefont {Li},\ and\ \citenamefont {Liu}}]{zhou2014epitaxial}%
	\BibitemOpen
	\bibfield  {author} {\bibinfo {author} {\bibfnamefont {M.}~\bibnamefont
			{Zhou}}, \bibinfo {author} {\bibfnamefont {W.}~\bibnamefont {Ming}}, \bibinfo
		{author} {\bibfnamefont {Z.}~\bibnamefont {Liu}}, \bibinfo {author}
		{\bibfnamefont {Z.}~\bibnamefont {Wang}}, \bibinfo {author} {\bibfnamefont
			{P.}~\bibnamefont {Li}}, \ and\ \bibinfo {author} {\bibfnamefont
			{F.}~\bibnamefont {Liu}},\ }\href@noop {} {\bibfield  {journal} {\bibinfo
			{journal} {Proceedings of the National Academy of Sciences}\ }\textbf
		{\bibinfo {volume} {111}},\ \bibinfo {pages} {14378} (\bibinfo {year}
		{2014})}\BibitemShut {NoStop}%
	\bibitem [{\citenamefont {Reis}\ \emph {et~al.}(2017)\citenamefont {Reis},
		\citenamefont {Li}, \citenamefont {Dudy}, \citenamefont {Bauernfeind},
		\citenamefont {Glass}, \citenamefont {Hanke}, \citenamefont {Thomale},
		\citenamefont {Sch{\"a}fer},\ and\ \citenamefont
		{Claessen}}]{reis2017bismuthene}%
	\BibitemOpen
	\bibfield  {author} {\bibinfo {author} {\bibfnamefont {F.}~\bibnamefont
			{Reis}}, \bibinfo {author} {\bibfnamefont {G.}~\bibnamefont {Li}}, \bibinfo
		{author} {\bibfnamefont {L.}~\bibnamefont {Dudy}}, \bibinfo {author}
		{\bibfnamefont {M.}~\bibnamefont {Bauernfeind}}, \bibinfo {author}
		{\bibfnamefont {S.}~\bibnamefont {Glass}}, \bibinfo {author} {\bibfnamefont
			{W.}~\bibnamefont {Hanke}}, \bibinfo {author} {\bibfnamefont
			{R.}~\bibnamefont {Thomale}}, \bibinfo {author} {\bibfnamefont
			{J.}~\bibnamefont {Sch{\"a}fer}}, \ and\ \bibinfo {author} {\bibfnamefont
			{R.}~\bibnamefont {Claessen}},\ }\href@noop {} {\bibfield  {journal}
		{\bibinfo  {journal} {Science}\ }\textbf {\bibinfo {volume} {357}},\ \bibinfo
		{pages} {287} (\bibinfo {year} {2017})}\BibitemShut {NoStop}%
	\bibitem [{\citenamefont {Wan}\ \emph {et~al.}(2011)\citenamefont {Wan},
		\citenamefont {Turner}, \citenamefont {Vishwanath},\ and\ \citenamefont
		{Savrasov}}]{wan2011topological}%
	\BibitemOpen
	\bibfield  {author} {\bibinfo {author} {\bibfnamefont {X.}~\bibnamefont
			{Wan}}, \bibinfo {author} {\bibfnamefont {A.~M.}\ \bibnamefont {Turner}},
		\bibinfo {author} {\bibfnamefont {A.}~\bibnamefont {Vishwanath}}, \ and\
		\bibinfo {author} {\bibfnamefont {S.~Y.}\ \bibnamefont {Savrasov}},\
	}\href@noop {} {\bibfield  {journal} {\bibinfo  {journal} {Physical Review
				B}\ }\textbf {\bibinfo {volume} {83}},\ \bibinfo {pages} {205101} (\bibinfo
		{year} {2011})}\BibitemShut {NoStop}%
	\bibitem [{\citenamefont {Xu}\ \emph {et~al.}(2015)\citenamefont {Xu},
		\citenamefont {Belopolski}, \citenamefont {Alidoust}, \citenamefont
		{Neupane}, \citenamefont {Bian}, \citenamefont {Zhang}, \citenamefont
		{Sankar}, \citenamefont {Chang}, \citenamefont {Yuan}, \citenamefont {Lee}
		\emph {et~al.}}]{xu2015discovery}%
	\BibitemOpen
	\bibfield  {author} {\bibinfo {author} {\bibfnamefont {S.-Y.}\ \bibnamefont
			{Xu}}, \bibinfo {author} {\bibfnamefont {I.}~\bibnamefont {Belopolski}},
		\bibinfo {author} {\bibfnamefont {N.}~\bibnamefont {Alidoust}}, \bibinfo
		{author} {\bibfnamefont {M.}~\bibnamefont {Neupane}}, \bibinfo {author}
		{\bibfnamefont {G.}~\bibnamefont {Bian}}, \bibinfo {author} {\bibfnamefont
			{C.}~\bibnamefont {Zhang}}, \bibinfo {author} {\bibfnamefont
			{R.}~\bibnamefont {Sankar}}, \bibinfo {author} {\bibfnamefont
			{G.}~\bibnamefont {Chang}}, \bibinfo {author} {\bibfnamefont
			{Z.}~\bibnamefont {Yuan}}, \bibinfo {author} {\bibfnamefont {C.-C.}\
			\bibnamefont {Lee}},  \emph {et~al.},\ }\href@noop {} {\bibfield  {journal}
		{\bibinfo  {journal} {Science}\ }\textbf {\bibinfo {volume} {349}},\ \bibinfo
		{pages} {613} (\bibinfo {year} {2015})}\BibitemShut {NoStop}%
	\bibitem [{\citenamefont {Lv}\ \emph {et~al.}(2015)\citenamefont {Lv},
		\citenamefont {Xu}, \citenamefont {Weng}, \citenamefont {Ma}, \citenamefont
		{Richard}, \citenamefont {Huang}, \citenamefont {Zhao}, \citenamefont {Chen},
		\citenamefont {Matt}, \citenamefont {Bisti} \emph
		{et~al.}}]{lv2015observation}%
	\BibitemOpen
	\bibfield  {author} {\bibinfo {author} {\bibfnamefont {B.}~\bibnamefont
			{Lv}}, \bibinfo {author} {\bibfnamefont {N.}~\bibnamefont {Xu}}, \bibinfo
		{author} {\bibfnamefont {H.}~\bibnamefont {Weng}}, \bibinfo {author}
		{\bibfnamefont {J.}~\bibnamefont {Ma}}, \bibinfo {author} {\bibfnamefont
			{P.}~\bibnamefont {Richard}}, \bibinfo {author} {\bibfnamefont
			{X.}~\bibnamefont {Huang}}, \bibinfo {author} {\bibfnamefont
			{L.}~\bibnamefont {Zhao}}, \bibinfo {author} {\bibfnamefont {G.}~\bibnamefont
			{Chen}}, \bibinfo {author} {\bibfnamefont {C.}~\bibnamefont {Matt}}, \bibinfo
		{author} {\bibfnamefont {F.}~\bibnamefont {Bisti}},  \emph {et~al.},\
	}\href@noop {} {\bibfield  {journal} {\bibinfo  {journal} {Nature Physics}\
		}\textbf {\bibinfo {volume} {11}},\ \bibinfo {pages} {724} (\bibinfo {year}
		{2015})}\BibitemShut {NoStop}%
	\bibitem [{\citenamefont {Bergholtz}\ \emph {et~al.}(2015)\citenamefont
		{Bergholtz}, \citenamefont {Liu}, \citenamefont {Trescher}, \citenamefont
		{Moessner},\ and\ \citenamefont {Udagawa}}]{bergholtz2015topology}%
	\BibitemOpen
	\bibfield  {author} {\bibinfo {author} {\bibfnamefont {E.~J.}\ \bibnamefont
			{Bergholtz}}, \bibinfo {author} {\bibfnamefont {Z.}~\bibnamefont {Liu}},
		\bibinfo {author} {\bibfnamefont {M.}~\bibnamefont {Trescher}}, \bibinfo
		{author} {\bibfnamefont {R.}~\bibnamefont {Moessner}}, \ and\ \bibinfo
		{author} {\bibfnamefont {M.}~\bibnamefont {Udagawa}},\ }\href@noop {}
	{\bibfield  {journal} {\bibinfo  {journal} {Physical review letters}\
		}\textbf {\bibinfo {volume} {114}},\ \bibinfo {pages} {016806} (\bibinfo
		{year} {2015})}\BibitemShut {NoStop}%
	\bibitem [{\citenamefont {Xu}\ \emph {et~al.}(2011)\citenamefont {Xu},
		\citenamefont {Weng}, \citenamefont {Wang}, \citenamefont {Dai},\ and\
		\citenamefont {Fang}}]{xu2011chern}%
	\BibitemOpen
	\bibfield  {author} {\bibinfo {author} {\bibfnamefont {G.}~\bibnamefont
			{Xu}}, \bibinfo {author} {\bibfnamefont {H.}~\bibnamefont {Weng}}, \bibinfo
		{author} {\bibfnamefont {Z.}~\bibnamefont {Wang}}, \bibinfo {author}
		{\bibfnamefont {X.}~\bibnamefont {Dai}}, \ and\ \bibinfo {author}
		{\bibfnamefont {Z.}~\bibnamefont {Fang}},\ }\href@noop {} {\bibfield
		{journal} {\bibinfo  {journal} {Physical review letters}\ }\textbf {\bibinfo
			{volume} {107}},\ \bibinfo {pages} {186806} (\bibinfo {year}
		{2011})}\BibitemShut {NoStop}%
	\bibitem [{\citenamefont {Yang}\ \emph {et~al.}(2015)\citenamefont {Yang},
		\citenamefont {Liu}, \citenamefont {Sun}, \citenamefont {Peng}, \citenamefont
		{Yang}, \citenamefont {Zhang}, \citenamefont {Zhou}, \citenamefont {Zhang},
		\citenamefont {Guo}, \citenamefont {Rahn} \emph {et~al.}}]{yang2015weyl}%
	\BibitemOpen
	\bibfield  {author} {\bibinfo {author} {\bibfnamefont {L.}~\bibnamefont
			{Yang}}, \bibinfo {author} {\bibfnamefont {Z.}~\bibnamefont {Liu}}, \bibinfo
		{author} {\bibfnamefont {Y.}~\bibnamefont {Sun}}, \bibinfo {author}
		{\bibfnamefont {H.}~\bibnamefont {Peng}}, \bibinfo {author} {\bibfnamefont
			{H.}~\bibnamefont {Yang}}, \bibinfo {author} {\bibfnamefont {T.}~\bibnamefont
			{Zhang}}, \bibinfo {author} {\bibfnamefont {B.}~\bibnamefont {Zhou}},
		\bibinfo {author} {\bibfnamefont {Y.}~\bibnamefont {Zhang}}, \bibinfo
		{author} {\bibfnamefont {Y.}~\bibnamefont {Guo}}, \bibinfo {author}
		{\bibfnamefont {M.}~\bibnamefont {Rahn}},  \emph {et~al.},\ }\href@noop {}
	{\bibfield  {journal} {\bibinfo  {journal} {Nature physics}\ }\textbf
		{\bibinfo {volume} {11}},\ \bibinfo {pages} {728} (\bibinfo {year}
		{2015})}\BibitemShut {NoStop}%
	\bibitem [{\citenamefont {Yan}\ and\ \citenamefont
		{Felser}(2017)}]{yan2017topological}%
	\BibitemOpen
	\bibfield  {author} {\bibinfo {author} {\bibfnamefont {B.}~\bibnamefont
			{Yan}}\ and\ \bibinfo {author} {\bibfnamefont {C.}~\bibnamefont {Felser}},\
	}\href@noop {} {\bibfield  {journal} {\bibinfo  {journal} {Annual Review of
				Condensed Matter Physics}\ }\textbf {\bibinfo {volume} {8}},\ \bibinfo
		{pages} {337} (\bibinfo {year} {2017})}\BibitemShut {NoStop}%
	\bibitem [{\citenamefont {Chang}\ \emph {et~al.}(2013)\citenamefont {Chang},
		\citenamefont {Zhang}, \citenamefont {Feng}, \citenamefont {Shen},
		\citenamefont {Zhang}, \citenamefont {Guo}, \citenamefont {Li}, \citenamefont
		{Ou}, \citenamefont {Wei}, \citenamefont {Wang} \emph
		{et~al.}}]{chang2013experimental}%
	\BibitemOpen
	\bibfield  {author} {\bibinfo {author} {\bibfnamefont {C.-Z.}\ \bibnamefont
			{Chang}}, \bibinfo {author} {\bibfnamefont {J.}~\bibnamefont {Zhang}},
		\bibinfo {author} {\bibfnamefont {X.}~\bibnamefont {Feng}}, \bibinfo {author}
		{\bibfnamefont {J.}~\bibnamefont {Shen}}, \bibinfo {author} {\bibfnamefont
			{Z.}~\bibnamefont {Zhang}}, \bibinfo {author} {\bibfnamefont
			{M.}~\bibnamefont {Guo}}, \bibinfo {author} {\bibfnamefont {K.}~\bibnamefont
			{Li}}, \bibinfo {author} {\bibfnamefont {Y.}~\bibnamefont {Ou}}, \bibinfo
		{author} {\bibfnamefont {P.}~\bibnamefont {Wei}}, \bibinfo {author}
		{\bibfnamefont {L.-L.}\ \bibnamefont {Wang}},  \emph {et~al.},\ }\href@noop
	{} {\bibfield  {journal} {\bibinfo  {journal} {Science}\ }\textbf {\bibinfo
			{volume} {340}},\ \bibinfo {pages} {167} (\bibinfo {year}
		{2013})}\BibitemShut {NoStop}%
	\bibitem [{\citenamefont {Yu}\ \emph {et~al.}(2010)\citenamefont {Yu},
		\citenamefont {Zhang}, \citenamefont {Zhang}, \citenamefont {Zhang},
		\citenamefont {Dai},\ and\ \citenamefont {Fang}}]{yu2010quantized}%
	\BibitemOpen
	\bibfield  {author} {\bibinfo {author} {\bibfnamefont {R.}~\bibnamefont
			{Yu}}, \bibinfo {author} {\bibfnamefont {W.}~\bibnamefont {Zhang}}, \bibinfo
		{author} {\bibfnamefont {H.-J.}\ \bibnamefont {Zhang}}, \bibinfo {author}
		{\bibfnamefont {S.-C.}\ \bibnamefont {Zhang}}, \bibinfo {author}
		{\bibfnamefont {X.}~\bibnamefont {Dai}}, \ and\ \bibinfo {author}
		{\bibfnamefont {Z.}~\bibnamefont {Fang}},\ }\href@noop {} {\bibfield
		{journal} {\bibinfo  {journal} {Science}\ }\textbf {\bibinfo {volume}
			{329}},\ \bibinfo {pages} {61} (\bibinfo {year} {2010})}\BibitemShut
	{NoStop}%
	\bibitem [{\citenamefont {Wang}\ \emph
		{et~al.}(2013{\natexlab{a}})\citenamefont {Wang}, \citenamefont {Liu},\ and\
		\citenamefont {Liu}}]{wang2013quantum}%
	\BibitemOpen
	\bibfield  {author} {\bibinfo {author} {\bibfnamefont {Z.}~\bibnamefont
			{Wang}}, \bibinfo {author} {\bibfnamefont {Z.}~\bibnamefont {Liu}}, \ and\
		\bibinfo {author} {\bibfnamefont {F.}~\bibnamefont {Liu}},\ }\href@noop {}
	{\bibfield  {journal} {\bibinfo  {journal} {Physical review letters}\
		}\textbf {\bibinfo {volume} {110}},\ \bibinfo {pages} {196801} (\bibinfo
		{year} {2013}{\natexlab{a}})}\BibitemShut {NoStop}%
	\bibitem [{\citenamefont {Liu}\ \emph {et~al.}(2014)\citenamefont {Liu},
		\citenamefont {Liu},\ and\ \citenamefont {Wu}}]{zheng2014exotic}%
	\BibitemOpen
	\bibfield  {author} {\bibinfo {author} {\bibfnamefont {Z.}~\bibnamefont
			{Liu}}, \bibinfo {author} {\bibfnamefont {F.}~\bibnamefont {Liu}}, \ and\
		\bibinfo {author} {\bibfnamefont {Y.-S.}\ \bibnamefont {Wu}},\ }\href@noop {}
	{\bibfield  {journal} {\bibinfo  {journal} {Chinese Physics B}\ }\textbf
		{\bibinfo {volume} {23}},\ \bibinfo {pages} {077308} (\bibinfo {year}
		{2014})}\BibitemShut {NoStop}%
	\bibitem [{\citenamefont
		{Mielke}(1991{\natexlab{a}})}]{mielke1991ferromagnetism}%
	\BibitemOpen
	\bibfield  {author} {\bibinfo {author} {\bibfnamefont {A.}~\bibnamefont
			{Mielke}},\ }\href@noop {} {\bibfield  {journal} {\bibinfo  {journal}
			{Journal of Physics A: Mathematical and General}\ }\textbf {\bibinfo {volume}
			{24}},\ \bibinfo {pages} {3311} (\bibinfo {year}
		{1991}{\natexlab{a}})}\BibitemShut {NoStop}%
	\bibitem [{\citenamefont
		{Mielke}(1991{\natexlab{b}})}]{mielke1991ferromagnetic}%
	\BibitemOpen
	\bibfield  {author} {\bibinfo {author} {\bibfnamefont {A.}~\bibnamefont
			{Mielke}},\ }\href@noop {} {\bibfield  {journal} {\bibinfo  {journal}
			{Journal of Physics A: Mathematical and General}\ }\textbf {\bibinfo {volume}
			{24}},\ \bibinfo {pages} {L73} (\bibinfo {year}
		{1991}{\natexlab{b}})}\BibitemShut {NoStop}%
	\bibitem [{\citenamefont {Mielke}(1992)}]{mielke1992exact}%
	\BibitemOpen
	\bibfield  {author} {\bibinfo {author} {\bibfnamefont {A.}~\bibnamefont
			{Mielke}},\ }\href@noop {} {\bibfield  {journal} {\bibinfo  {journal}
			{Journal of Physics A: Mathematical and General}\ }\textbf {\bibinfo {volume}
			{25}},\ \bibinfo {pages} {4335} (\bibinfo {year} {1992})}\BibitemShut
	{NoStop}%
	\bibitem [{\citenamefont {Tasaki}(1992)}]{tasaki1992ferromagnetism}%
	\BibitemOpen
	\bibfield  {author} {\bibinfo {author} {\bibfnamefont {H.}~\bibnamefont
			{Tasaki}},\ }\href@noop {} {\bibfield  {journal} {\bibinfo  {journal}
			{Physical review letters}\ }\textbf {\bibinfo {volume} {69}},\ \bibinfo
		{pages} {1608} (\bibinfo {year} {1992})}\BibitemShut {NoStop}%
	\bibitem [{\citenamefont {Zhang}\ \emph {et~al.}(2010)\citenamefont {Zhang},
		\citenamefont {Hung},\ and\ \citenamefont {Wu}}]{zhang2010proposed}%
	\BibitemOpen
	\bibfield  {author} {\bibinfo {author} {\bibfnamefont {S.}~\bibnamefont
			{Zhang}}, \bibinfo {author} {\bibfnamefont {H.-h.}\ \bibnamefont {Hung}}, \
		and\ \bibinfo {author} {\bibfnamefont {C.}~\bibnamefont {Wu}},\ }\href@noop
	{} {\bibfield  {journal} {\bibinfo  {journal} {Physical Review A}\ }\textbf
		{\bibinfo {volume} {82}},\ \bibinfo {pages} {053618} (\bibinfo {year}
		{2010})}\BibitemShut {NoStop}%
	\bibitem [{\citenamefont {Wu}\ \emph {et~al.}(2007)\citenamefont {Wu},
		\citenamefont {Bergman}, \citenamefont {Balents},\ and\ \citenamefont
		{Sarma}}]{wu2007flat}%
	\BibitemOpen
	\bibfield  {author} {\bibinfo {author} {\bibfnamefont {C.}~\bibnamefont
			{Wu}}, \bibinfo {author} {\bibfnamefont {D.}~\bibnamefont {Bergman}},
		\bibinfo {author} {\bibfnamefont {L.}~\bibnamefont {Balents}}, \ and\
		\bibinfo {author} {\bibfnamefont {S.~D.}\ \bibnamefont {Sarma}},\ }\href@noop
	{} {\bibfield  {journal} {\bibinfo  {journal} {Physical review letters}\
		}\textbf {\bibinfo {volume} {99}},\ \bibinfo {pages} {070401} (\bibinfo
		{year} {2007})}\BibitemShut {NoStop}%
	\bibitem [{\citenamefont {Wu}\ and\ \citenamefont {Sarma}(2008)}]{wu2008p}%
	\BibitemOpen
	\bibfield  {author} {\bibinfo {author} {\bibfnamefont {C.}~\bibnamefont
			{Wu}}\ and\ \bibinfo {author} {\bibfnamefont {S.~D.}\ \bibnamefont {Sarma}},\
	}\href@noop {} {\bibfield  {journal} {\bibinfo  {journal} {Physical Review
				B}\ }\textbf {\bibinfo {volume} {77}},\ \bibinfo {pages} {235107} (\bibinfo
		{year} {2008})}\BibitemShut {NoStop}%
	\bibitem [{\citenamefont {Miyahara}\ \emph {et~al.}(2007)\citenamefont
		{Miyahara}, \citenamefont {Kusuta},\ and\ \citenamefont
		{Furukawa}}]{miyahara2007bcs}%
	\BibitemOpen
	\bibfield  {author} {\bibinfo {author} {\bibfnamefont {S.}~\bibnamefont
			{Miyahara}}, \bibinfo {author} {\bibfnamefont {S.}~\bibnamefont {Kusuta}}, \
		and\ \bibinfo {author} {\bibfnamefont {N.}~\bibnamefont {Furukawa}},\
	}\href@noop {} {\bibfield  {journal} {\bibinfo  {journal} {Physica C:
				Superconductivity}\ }\textbf {\bibinfo {volume} {460}},\ \bibinfo {pages}
		{1145} (\bibinfo {year} {2007})}\BibitemShut {NoStop}%
	\bibitem [{\citenamefont {Kobayashi}\ \emph {et~al.}(2016)\citenamefont
		{Kobayashi}, \citenamefont {Okumura}, \citenamefont {Yamada}, \citenamefont
		{Machida},\ and\ \citenamefont {Aoki}}]{kobayashi2016superconductivity}%
	\BibitemOpen
	\bibfield  {author} {\bibinfo {author} {\bibfnamefont {K.}~\bibnamefont
			{Kobayashi}}, \bibinfo {author} {\bibfnamefont {M.}~\bibnamefont {Okumura}},
		\bibinfo {author} {\bibfnamefont {S.}~\bibnamefont {Yamada}}, \bibinfo
		{author} {\bibfnamefont {M.}~\bibnamefont {Machida}}, \ and\ \bibinfo
		{author} {\bibfnamefont {H.}~\bibnamefont {Aoki}},\ }\href@noop {} {\bibfield
		{journal} {\bibinfo  {journal} {Physical Review B}\ }\textbf {\bibinfo
			{volume} {94}},\ \bibinfo {pages} {214501} (\bibinfo {year}
		{2016})}\BibitemShut {NoStop}%
	\bibitem [{\citenamefont {Tang}\ \emph {et~al.}(2011)\citenamefont {Tang},
		\citenamefont {Mei},\ and\ \citenamefont {Wen}}]{tang2011high}%
	\BibitemOpen
	\bibfield  {author} {\bibinfo {author} {\bibfnamefont {E.}~\bibnamefont
			{Tang}}, \bibinfo {author} {\bibfnamefont {J.-W.}\ \bibnamefont {Mei}}, \
		and\ \bibinfo {author} {\bibfnamefont {X.-G.}\ \bibnamefont {Wen}},\
	}\href@noop {} {\bibfield  {journal} {\bibinfo  {journal} {Physical review
				letters}\ }\textbf {\bibinfo {volume} {106}},\ \bibinfo {pages} {236802}
		(\bibinfo {year} {2011})}\BibitemShut {NoStop}%
	\bibitem [{\citenamefont {Neupert}\ \emph {et~al.}(2011)\citenamefont
		{Neupert}, \citenamefont {Santos}, \citenamefont {Chamon},\ and\
		\citenamefont {Mudry}}]{neupert2011fractional}%
	\BibitemOpen
	\bibfield  {author} {\bibinfo {author} {\bibfnamefont {T.}~\bibnamefont
			{Neupert}}, \bibinfo {author} {\bibfnamefont {L.}~\bibnamefont {Santos}},
		\bibinfo {author} {\bibfnamefont {C.}~\bibnamefont {Chamon}}, \ and\ \bibinfo
		{author} {\bibfnamefont {C.}~\bibnamefont {Mudry}},\ }\href@noop {}
	{\bibfield  {journal} {\bibinfo  {journal} {Physical review letters}\
		}\textbf {\bibinfo {volume} {106}},\ \bibinfo {pages} {236804} (\bibinfo
		{year} {2011})}\BibitemShut {NoStop}%
	\bibitem [{\citenamefont {Kimura}\ \emph {et~al.}(2002)\citenamefont {Kimura},
		\citenamefont {Tamura}, \citenamefont {Shiraishi},\ and\ \citenamefont
		{Takayanagi}}]{kimura2002magnetic}%
	\BibitemOpen
	\bibfield  {author} {\bibinfo {author} {\bibfnamefont {T.}~\bibnamefont
			{Kimura}}, \bibinfo {author} {\bibfnamefont {H.}~\bibnamefont {Tamura}},
		\bibinfo {author} {\bibfnamefont {K.}~\bibnamefont {Shiraishi}}, \ and\
		\bibinfo {author} {\bibfnamefont {H.}~\bibnamefont {Takayanagi}},\
	}\href@noop {} {\bibfield  {journal} {\bibinfo  {journal} {Physical Review
				B}\ }\textbf {\bibinfo {volume} {65}},\ \bibinfo {pages} {081307} (\bibinfo
		{year} {2002})}\BibitemShut {NoStop}%
	\bibitem [{\citenamefont {Tanaka}\ and\ \citenamefont
		{Ueda}(2003)}]{tanaka2003stability}%
	\BibitemOpen
	\bibfield  {author} {\bibinfo {author} {\bibfnamefont {A.}~\bibnamefont
			{Tanaka}}\ and\ \bibinfo {author} {\bibfnamefont {H.}~\bibnamefont {Ueda}},\
	}\href@noop {} {\bibfield  {journal} {\bibinfo  {journal} {Physical review
				letters}\ }\textbf {\bibinfo {volume} {90}},\ \bibinfo {pages} {067204}
		(\bibinfo {year} {2003})}\BibitemShut {NoStop}%
	\bibitem [{\citenamefont {Pollmann}\ \emph {et~al.}(2008)\citenamefont
		{Pollmann}, \citenamefont {Fulde},\ and\ \citenamefont
		{Shtengel}}]{pollmann2008kinetic}%
	\BibitemOpen
	\bibfield  {author} {\bibinfo {author} {\bibfnamefont {F.}~\bibnamefont
			{Pollmann}}, \bibinfo {author} {\bibfnamefont {P.}~\bibnamefont {Fulde}}, \
		and\ \bibinfo {author} {\bibfnamefont {K.}~\bibnamefont {Shtengel}},\
	}\href@noop {} {\bibfield  {journal} {\bibinfo  {journal} {Physical review
				letters}\ }\textbf {\bibinfo {volume} {100}},\ \bibinfo {pages} {136404}
		(\bibinfo {year} {2008})}\BibitemShut {NoStop}%
	\bibitem [{\citenamefont {Hase}\ \emph {et~al.}(2018)\citenamefont {Hase},
		\citenamefont {Yanagisawa}, \citenamefont {Aiura},\ and\ \citenamefont
		{Kawashima}}]{hase2018possibility}%
	\BibitemOpen
	\bibfield  {author} {\bibinfo {author} {\bibfnamefont {I.}~\bibnamefont
			{Hase}}, \bibinfo {author} {\bibfnamefont {T.}~\bibnamefont {Yanagisawa}},
		\bibinfo {author} {\bibfnamefont {Y.}~\bibnamefont {Aiura}}, \ and\ \bibinfo
		{author} {\bibfnamefont {K.}~\bibnamefont {Kawashima}},\ }\href@noop {}
	{\bibfield  {journal} {\bibinfo  {journal} {Physical Review Letters}\
		}\textbf {\bibinfo {volume} {120}},\ \bibinfo {pages} {196401} (\bibinfo
		{year} {2018})}\BibitemShut {NoStop}%
	\bibitem [{\citenamefont {Liu}\ \emph {et~al.}(2009)\citenamefont {Liu},
		\citenamefont {Zhang}, \citenamefont {Wang},\ and\ \citenamefont
		{Li}}]{liu2009spin}%
	\BibitemOpen
	\bibfield  {author} {\bibinfo {author} {\bibfnamefont {G.}~\bibnamefont
			{Liu}}, \bibinfo {author} {\bibfnamefont {P.}~\bibnamefont {Zhang}}, \bibinfo
		{author} {\bibfnamefont {Z.}~\bibnamefont {Wang}}, \ and\ \bibinfo {author}
		{\bibfnamefont {S.-S.}\ \bibnamefont {Li}},\ }\href@noop {} {\bibfield
		{journal} {\bibinfo  {journal} {Physical Review B}\ }\textbf {\bibinfo
			{volume} {79}},\ \bibinfo {pages} {035323} (\bibinfo {year}
		{2009})}\BibitemShut {NoStop}%
	\bibitem [{\citenamefont {Guo}\ and\ \citenamefont
		{Franz}(2009{\natexlab{a}})}]{guo2009three}%
	\BibitemOpen
	\bibfield  {author} {\bibinfo {author} {\bibfnamefont {H.-M.}\ \bibnamefont
			{Guo}}\ and\ \bibinfo {author} {\bibfnamefont {M.}~\bibnamefont {Franz}},\
	}\href@noop {} {\bibfield  {journal} {\bibinfo  {journal} {Physical review
				letters}\ }\textbf {\bibinfo {volume} {103}},\ \bibinfo {pages} {206805}
		(\bibinfo {year} {2009}{\natexlab{a}})}\BibitemShut {NoStop}%
	\bibitem [{\citenamefont {Guo}\ and\ \citenamefont
		{Franz}(2009{\natexlab{b}})}]{guo2009topological}%
	\BibitemOpen
	\bibfield  {author} {\bibinfo {author} {\bibfnamefont {H.-M.}\ \bibnamefont
			{Guo}}\ and\ \bibinfo {author} {\bibfnamefont {M.}~\bibnamefont {Franz}},\
	}\href@noop {} {\bibfield  {journal} {\bibinfo  {journal} {Physical Review
				B}\ }\textbf {\bibinfo {volume} {80}},\ \bibinfo {pages} {113102} (\bibinfo
		{year} {2009}{\natexlab{b}})}\BibitemShut {NoStop}%
	\bibitem [{\citenamefont {Wang}\ and\ \citenamefont
		{Zhang}(2010)}]{wang2010quantum}%
	\BibitemOpen
	\bibfield  {author} {\bibinfo {author} {\bibfnamefont {Z.}~\bibnamefont
			{Wang}}\ and\ \bibinfo {author} {\bibfnamefont {P.}~\bibnamefont {Zhang}},\
	}\href@noop {} {\bibfield  {journal} {\bibinfo  {journal} {New Journal of
				Physics}\ }\textbf {\bibinfo {volume} {12}},\ \bibinfo {pages} {043055}
		(\bibinfo {year} {2010})}\BibitemShut {NoStop}%
	\bibitem [{\citenamefont {Kurita}\ \emph {et~al.}(2011)\citenamefont {Kurita},
		\citenamefont {Yamaji},\ and\ \citenamefont {Imada}}]{kurita2011topological}%
	\BibitemOpen
	\bibfield  {author} {\bibinfo {author} {\bibfnamefont {M.}~\bibnamefont
			{Kurita}}, \bibinfo {author} {\bibfnamefont {Y.}~\bibnamefont {Yamaji}}, \
		and\ \bibinfo {author} {\bibfnamefont {M.}~\bibnamefont {Imada}},\
	}\href@noop {} {\bibfield  {journal} {\bibinfo  {journal} {Journal of the
				Physical Society of Japan}\ }\textbf {\bibinfo {volume} {80}},\ \bibinfo
		{pages} {044708} (\bibinfo {year} {2011})}\BibitemShut {NoStop}%
	\bibitem [{\citenamefont {Hatsugai}\ and\ \citenamefont
		{Maruyama}(2011)}]{hatsugai2011zq}%
	\BibitemOpen
	\bibfield  {author} {\bibinfo {author} {\bibfnamefont {Y.}~\bibnamefont
			{Hatsugai}}\ and\ \bibinfo {author} {\bibfnamefont {I.}~\bibnamefont
			{Maruyama}},\ }\href@noop {} {\bibfield  {journal} {\bibinfo  {journal} {EPL
				(Europhysics Letters)}\ }\textbf {\bibinfo {volume} {95}},\ \bibinfo {pages}
		{20003} (\bibinfo {year} {2011})}\BibitemShut {NoStop}%
	\bibitem [{\citenamefont {Sun}\ \emph {et~al.}(2011)\citenamefont {Sun},
		\citenamefont {Gu}, \citenamefont {Katsura},\ and\ \citenamefont
		{Sarma}}]{sun2011nearly}%
	\BibitemOpen
	\bibfield  {author} {\bibinfo {author} {\bibfnamefont {K.}~\bibnamefont
			{Sun}}, \bibinfo {author} {\bibfnamefont {Z.}~\bibnamefont {Gu}}, \bibinfo
		{author} {\bibfnamefont {H.}~\bibnamefont {Katsura}}, \ and\ \bibinfo
		{author} {\bibfnamefont {S.~D.}\ \bibnamefont {Sarma}},\ }\href@noop {}
	{\bibfield  {journal} {\bibinfo  {journal} {Physical review letters}\
		}\textbf {\bibinfo {volume} {106}},\ \bibinfo {pages} {236803} (\bibinfo
		{year} {2011})}\BibitemShut {NoStop}%
	\bibitem [{\citenamefont {Wang}\ \emph
		{et~al.}(2013{\natexlab{b}})\citenamefont {Wang}, \citenamefont {Su},\ and\
		\citenamefont {Liu}}]{wang2013prediction}%
	\BibitemOpen
	\bibfield  {author} {\bibinfo {author} {\bibfnamefont {Z.}~\bibnamefont
			{Wang}}, \bibinfo {author} {\bibfnamefont {N.}~\bibnamefont {Su}}, \ and\
		\bibinfo {author} {\bibfnamefont {F.}~\bibnamefont {Liu}},\ }\href@noop {}
	{\bibfield  {journal} {\bibinfo  {journal} {Nano letters}\ }\textbf {\bibinfo
			{volume} {13}},\ \bibinfo {pages} {2842} (\bibinfo {year}
		{2013}{\natexlab{b}})}\BibitemShut {NoStop}%
	\bibitem [{Note1()}]{Note1}%
	\BibitemOpen
	\bibinfo {note} {See Supplemental Materials at
		http://link.aps.org/supplemental/xxx, for more details which include
		Ref.~\cite
		{blochl1994projector,kresse1999ultrasoft,perdew1996generalized,perdew1996generalized,kresse1996efficient,methfessel1989high,hobbs2000fully,marzari1997maximally,wu2018wanniertools}.}\BibitemShut
	{Stop}%
	\bibitem [{40f()}]{40footnote}%
	\BibitemOpen
	\href@noop {} {\bibinfo  {journal} {One might view the pyrochlore lattice as
			a 3D version of Kagome lattice, as shown in Fig. S2. Effectively one stacks
			the 2D Kagome layers in three orientations in a pyrochlore lattice. On the
			other hand, the pyrochlore lattice is a 3D line graph of diamond lattice, as
			shown in Fig. S3, which amounts to a 3D frustrated hopping model, giving rise
			to the FBs with destructive interference}\ }\BibitemShut {NoStop}%
	\bibitem [{\citenamefont {Thouless}\ \emph {et~al.}(1982)\citenamefont
		{Thouless}, \citenamefont {Kohmoto}, \citenamefont {Nightingale},\ and\
		\citenamefont {den Nijs}}]{thouless1982quantized}%
	\BibitemOpen
	\bibfield  {journal} {  }\bibfield  {author} {\bibinfo {author} {\bibfnamefont
			{D.~J.}\ \bibnamefont {Thouless}}, \bibinfo {author} {\bibfnamefont
			{M.}~\bibnamefont {Kohmoto}}, \bibinfo {author} {\bibfnamefont {M.~P.}\
			\bibnamefont {Nightingale}}, \ and\ \bibinfo {author} {\bibfnamefont
			{M.}~\bibnamefont {den Nijs}},\ }\href@noop {} {\bibfield  {journal}
		{\bibinfo  {journal} {Physical Review Letters}\ }\textbf {\bibinfo {volume}
			{49}},\ \bibinfo {pages} {405} (\bibinfo {year} {1982})}\BibitemShut
	{NoStop}%
	\bibitem [{\citenamefont {Kane}\ and\ \citenamefont
		{Mele}(2005{\natexlab{b}})}]{kane2005z}%
	\BibitemOpen
	\bibfield  {author} {\bibinfo {author} {\bibfnamefont {C.~L.}\ \bibnamefont
			{Kane}}\ and\ \bibinfo {author} {\bibfnamefont {E.~J.}\ \bibnamefont
			{Mele}},\ }\href@noop {} {\bibfield  {journal} {\bibinfo  {journal} {Physical
				review letters}\ }\textbf {\bibinfo {volume} {95}},\ \bibinfo {pages}
		{146802} (\bibinfo {year} {2005}{\natexlab{b}})}\BibitemShut {NoStop}%
	\bibitem [{\citenamefont {Mostofi}\ \emph {et~al.}(2008)\citenamefont
		{Mostofi}, \citenamefont {Yates}, \citenamefont {Lee}, \citenamefont {Souza},
		\citenamefont {Vanderbilt},\ and\ \citenamefont
		{Marzari}}]{mostofi2008wannier90}%
	\BibitemOpen
	\bibfield  {author} {\bibinfo {author} {\bibfnamefont {A.~A.}\ \bibnamefont
			{Mostofi}}, \bibinfo {author} {\bibfnamefont {J.~R.}\ \bibnamefont {Yates}},
		\bibinfo {author} {\bibfnamefont {Y.-S.}\ \bibnamefont {Lee}}, \bibinfo
		{author} {\bibfnamefont {I.}~\bibnamefont {Souza}}, \bibinfo {author}
		{\bibfnamefont {D.}~\bibnamefont {Vanderbilt}}, \ and\ \bibinfo {author}
		{\bibfnamefont {N.}~\bibnamefont {Marzari}},\ }\href@noop {} {\bibfield
		{journal} {\bibinfo  {journal} {Computer physics communications}\ }\textbf
		{\bibinfo {volume} {178}},\ \bibinfo {pages} {685} (\bibinfo {year}
		{2008})}\BibitemShut {NoStop}%
	\bibitem [{\citenamefont {Bl{\"o}chl}(1994)}]{blochl1994projector}%
	\BibitemOpen
	\bibfield  {author} {\bibinfo {author} {\bibfnamefont {P.~E.}\ \bibnamefont
			{Bl{\"o}chl}},\ }\href@noop {} {\bibfield  {journal} {\bibinfo  {journal}
			{Physical review B}\ }\textbf {\bibinfo {volume} {50}},\ \bibinfo {pages}
		{17953} (\bibinfo {year} {1994})}\BibitemShut {NoStop}%
	\bibitem [{\citenamefont {Kresse}\ and\ \citenamefont
		{Joubert}(1999)}]{kresse1999ultrasoft}%
	\BibitemOpen
	\bibfield  {author} {\bibinfo {author} {\bibfnamefont {G.}~\bibnamefont
			{Kresse}}\ and\ \bibinfo {author} {\bibfnamefont {D.}~\bibnamefont
			{Joubert}},\ }\href@noop {} {\bibfield  {journal} {\bibinfo  {journal}
			{Physical Review B}\ }\textbf {\bibinfo {volume} {59}},\ \bibinfo {pages}
		{1758} (\bibinfo {year} {1999})}\BibitemShut {NoStop}%
	\bibitem [{\citenamefont {Perdew}\ \emph {et~al.}(1996)\citenamefont {Perdew},
		\citenamefont {Burke},\ and\ \citenamefont
		{Ernzerhof}}]{perdew1996generalized}%
	\BibitemOpen
	\bibfield  {author} {\bibinfo {author} {\bibfnamefont {J.~P.}\ \bibnamefont
			{Perdew}}, \bibinfo {author} {\bibfnamefont {K.}~\bibnamefont {Burke}}, \
		and\ \bibinfo {author} {\bibfnamefont {M.}~\bibnamefont {Ernzerhof}},\
	}\href@noop {} {\bibfield  {journal} {\bibinfo  {journal} {Physical review
				letters}\ }\textbf {\bibinfo {volume} {77}},\ \bibinfo {pages} {3865}
		(\bibinfo {year} {1996})}\BibitemShut {NoStop}%
	\bibitem [{\citenamefont {Kresse}\ and\ \citenamefont
		{Furthm{\"u}ller}(1996)}]{kresse1996efficient}%
	\BibitemOpen
	\bibfield  {author} {\bibinfo {author} {\bibfnamefont {G.}~\bibnamefont
			{Kresse}}\ and\ \bibinfo {author} {\bibfnamefont {J.}~\bibnamefont
			{Furthm{\"u}ller}},\ }\href@noop {} {\bibfield  {journal} {\bibinfo
			{journal} {Physical review B}\ }\textbf {\bibinfo {volume} {54}},\ \bibinfo
		{pages} {11169} (\bibinfo {year} {1996})}\BibitemShut {NoStop}%
	\bibitem [{\citenamefont {Methfessel}\ and\ \citenamefont
		{Paxton}(1989)}]{methfessel1989high}%
	\BibitemOpen
	\bibfield  {author} {\bibinfo {author} {\bibfnamefont {M.}~\bibnamefont
			{Methfessel}}\ and\ \bibinfo {author} {\bibfnamefont {A.}~\bibnamefont
			{Paxton}},\ }\href@noop {} {\bibfield  {journal} {\bibinfo  {journal}
			{Physical Review B}\ }\textbf {\bibinfo {volume} {40}},\ \bibinfo {pages}
		{3616} (\bibinfo {year} {1989})}\BibitemShut {NoStop}%
	\bibitem [{\citenamefont {Hobbs}\ \emph {et~al.}(2000)\citenamefont {Hobbs},
		\citenamefont {Kresse},\ and\ \citenamefont {Hafner}}]{hobbs2000fully}%
	\BibitemOpen
	\bibfield  {author} {\bibinfo {author} {\bibfnamefont {D.}~\bibnamefont
			{Hobbs}}, \bibinfo {author} {\bibfnamefont {G.}~\bibnamefont {Kresse}}, \
		and\ \bibinfo {author} {\bibfnamefont {J.}~\bibnamefont {Hafner}},\
	}\href@noop {} {\bibfield  {journal} {\bibinfo  {journal} {Physical Review
				B}\ }\textbf {\bibinfo {volume} {62}},\ \bibinfo {pages} {11556} (\bibinfo
		{year} {2000})}\BibitemShut {NoStop}%
	\bibitem [{\citenamefont {Marzari}\ and\ \citenamefont
		{Vanderbilt}(1997)}]{marzari1997maximally}%
	\BibitemOpen
	\bibfield  {author} {\bibinfo {author} {\bibfnamefont {N.}~\bibnamefont
			{Marzari}}\ and\ \bibinfo {author} {\bibfnamefont {D.}~\bibnamefont
			{Vanderbilt}},\ }\href@noop {} {\bibfield  {journal} {\bibinfo  {journal}
			{Physical review B}\ }\textbf {\bibinfo {volume} {56}},\ \bibinfo {pages}
		{12847} (\bibinfo {year} {1997})}\BibitemShut {NoStop}%
	\bibitem [{\citenamefont {Wu}\ \emph {et~al.}(2018)\citenamefont {Wu},
		\citenamefont {Zhang}, \citenamefont {Song}, \citenamefont {Troyer},\ and\
		\citenamefont {Soluyanov}}]{wu2018wanniertools}%
	\BibitemOpen
	\bibfield  {author} {\bibinfo {author} {\bibfnamefont {Q.}~\bibnamefont
			{Wu}}, \bibinfo {author} {\bibfnamefont {S.}~\bibnamefont {Zhang}}, \bibinfo
		{author} {\bibfnamefont {H.-F.}\ \bibnamefont {Song}}, \bibinfo {author}
		{\bibfnamefont {M.}~\bibnamefont {Troyer}}, \ and\ \bibinfo {author}
		{\bibfnamefont {A.~A.}\ \bibnamefont {Soluyanov}},\ }\href@noop {} {\bibfield
		{journal} {\bibinfo  {journal} {Computer Physics Communications}\ }\textbf
		{\bibinfo {volume} {224}},\ \bibinfo {pages} {405} (\bibinfo {year}
		{2018})}\BibitemShut {NoStop}%
\end{thebibliography}

\providecommand{\noopsort}[1]{}\providecommand{\singleletter}[1]{#1}%

\end{document}